\documentclass{article}

\usepackage{arxiv}

\usepackage{amssymb}
\usepackage{subfigure}
\usepackage{color} %red font
\usepackage{amsmath}
\usepackage{url}
\usepackage{algorithm}
\usepackage{algorithmicx}
\usepackage{algpseudocode}
\usepackage{diagbox} %table
\usepackage{booktabs} %table
\usepackage[figuresright]{rotating}

\title{CasSeqGCN: Combining Network Structure and Temporal Sequence to Predict Information Cascades}

\author{
 Yansong Wang \\
  College of Computer and Information Science\\
  Southwest University\\
  Chongqing, 400715, P. R. China \\
  \texttt{yansong0682@email.swu.edu.cn} \\
   \And
 Xiaomeng Wang \\
  College of Computer and Information Science\\
  Southwest University\\
  Chongqing, 400715, P. R. China \\
  \texttt{wxm1706@swu.edu.cn} \\
  \And
 Yijun Ran \\
  College of Computer and Information Science\\
  Southwest University\\
  Chongqing, 400715, P. R. China \\
  \texttt{ranyij288@email.swu.edu.cn} \\
  \And
  Radosław Michalski \\
  Faculty of Computer Science and Management\\
  Wrocław University of Science and Technology, Poland\\
  \texttt{radoslaw.michalski@pwr.edu.pl} \\
  \And
  Tao Jia \\
  College of Computer and Information Science\\
  Southwest University\\
  Chongqing, 400715, P. R. China \\
  \texttt{tjia@swu.edu.cn, corresponding author} \\
}

\begin{document}
\maketitle
\begin{abstract}
One important task in the study of information cascade is to predict the future recipients of a message given its past spreading trajectory. While the network structure serves as the backbone of the spreading, an accurate prediction can hardly be made without the knowledge of the dynamics on the network. The temporal information in the spreading sequence captures many hidden features, but predictions based on sequence alone have their limitations. Recent efforts start to explore the possibility of combining both the network structure and the temporal feature. Here, we propose a new end-to-end prediction method CasSeqGCN in which the structure and temporal feature are simultaneously taken into account. A cascade is divided into multiple snapshots which record the network topology and the state of nodes. The graph convolutional network (GCN) is used to learn the representation of a snapshot. A novel aggregation method based on dynamic routing is proposed to aggregate node representation and the long short-term memory (LSTM) model is used to extract temporal information. CasSeqGCN predicts the future cascade size more accurately compared with other state-of-art baseline methods. The ablation study demonstrates that the improvement mainly comes from the design of the input and the GCN layer. We explicitly design an experiment to show the quality of the cascade representation learned by our approach is better than other methods. Our work proposes a new approach to combine the structural and temporal features, which not only gives a useful baseline model for future studies of cascade prediction, but also brings new insights on a wide collection of problems related with dynamics on and of the network.
\end{abstract}

% keywords can be removed

\section{Introduction}
Online social networks, such as Facebook, Twitter, Weibo and Youtube, play an increasingly important role in our daily life. They not only change how we communicate and interact with each other, but also provide large platforms for social computing studies \cite{watts2016computational}. Understanding how information propagates in the social network is a direction that draws wide attentions, which includes diverse topics such as the dynamics of the spreading \cite{zhang2016dynamics}, the underlying mechanism \cite{centola2010spread}, the role of the network topology and more \cite{xie2021detecting,jia2015analysis,kianian2021efficient,su2020emergence}. Cascade prediction is a task that is interested by computer scientists\cite{cheng2014can, zhou2021survey}, with important applications to several real-world problems such as viral marketing \cite{robles2020evolutionary}, recommendation \cite{wu2019dual} and information security \cite{zhao2020online}.

The task of cascade prediction is to use the information of the observed spreading to make a prediction that best matches with the future spreading. One approach is to model the dynamical process of information spreading on networks. If the pattern can be generalized into a simple model with a few parameters, and the topology of the underlying network is known, one can fit the model with observations to forecast the trends in the future. The network structure definitely provides the backbone of the spreading. However, it is only one of many factors that determine what a real spreading sequence would look like. The burst in our human activity, the social reinforcement effect, the individual difference on the path of the spreading, and the missing or unknown links in the given network would all affect the final outcome, which can not be captured by a simple model \cite{cheng2014can}. Another approach is to use the traditional machine learning method, where we can throw all features that are thought to be important in a predictor \cite{tsur2012s,ma2013predicting,cui2013cascading,weng2014predicting}. But the problem is that we have to first select and quantify the feature we are interested in. The number of nearest neighbors, the number of the second and the third nearest neighbors are all related to neighborhood information, but this information alone has multiple or even infinite numbers of ways to quantify. Thanks to the deep learning method, an end-to-end prediction is now possible, in which one does not need to worry about picking a feature and its exact expression. A lot of works consider using the network structure and temporal dynamics as the input, as they are independent of the social network platform and the information content, and can be generalized to different circumstances \cite{li2017deepcas,chen2019information}. Recent efforts start to combine both of them to enhance the quality of prediction. Such approaches utilize the network embedding and temporal dynamics with further applications to a wide range of problems such as traffic forecasting and anomaly detection \cite{zhao2019t,bai2021a3t,zheng2019addgraph}.

In this paper, we propose CasSeqGCN, an end-to-end deep learning framework, which combines the structural and temporal features to predict the growth size of a given cascade. By defining the active/inactive state of a node, we divide one cascade into multiple snapshots, each containing the topology of the underlying network and the status of whether the information has reached a node. The representation of nodes in one snapshot are learned by the classical graph convolutional network (GCN) \cite{kipf2016semi}. The output embedding of all nodes is further aggregated into one vector by dynamic routing, forming a representation of one snapshot. A sequence of the snapshot, with the structure information embedded, is sent to the long short-term memory (LSTM) \cite{hochreiter1997long} layer to extract the temporal order of spreading. The CasSeqGCN gives much better prediction on three distinct datasets compared with several state-of-art models. We perform extended ablation studies to exam the contribution of the enhancement from different parts of CasSeqGCN model. While aggregation and temporal feature learning are both important, the way the input is designed and the use of GCN layer is found to be most crucial to the improvement. The learned representation well captures the structural and temporal information in a cascade, which demonstrates potential in other tasks. The CasSeqGCN provides new insights on how to learn and combine structural and temporal information for prediction, as well as a new baseline for future studies on cascade prediction.

The contribution of our work can be summarized as follows. First, we introduce a new approach to combine the structural and temporal features that can be applied to general studies related to dynamics of and on the network. The approach is simple to implement with low computational cost. Second, the CasSeqGCN model for cascade prediction outperforms other existing models, which can serve as a new baseline for future studies. The code is publicly available at \url{https://github.com/MrYansong/CasSeqGCN} for future reference and reproducibility. Finally, we explicitly illustrate the contribution of each component in our model, which sheds light on future modifications and improvements. We particularly design an experiment to test the quality of cascade representations learned by CasSeqGCN and other methods. For the same training data and classifier, the input cascade representation by CasSeqGCN gives the best performance, suggesting that our approach has the capability to accurately learn the representation of the cascade. In the following, we first briefly review related works in section 2. The detailed introduction of the CasSeqGCN model is presented in section 3. In section 4, we introduce the description of the datasets, the baseline methods, the setup and the results of the experiment. We also perform ablation studies in this section to separately analyze the contribution of each part of the model. In section 5, we make a short summary and description for future work.

\section{Related Work}
Similar to many tasks in computer science, prediction of information cascade can be generally divided into two categories: classification and regression \cite{zhou2021survey}. 
Classification \cite{weng2014predicting,shulman2016predictability,kefato2018cas2vec,liao2019popularity} involves the prediction on whether the number of retweets of a message will exceed a certain threshold \cite{gou2018learning}, or fall into one of the predefined popularity ranges \cite{hong2011predicting,liao2019popularity}. The regression problem \cite{kupavskii2013predicting,li2017deepcas,cao2017deephawkes,chen2019information} is generally more complicated than classification, which usually aims to predict the future spreading sequence or spreading size using information from the observation such as network structure, user attributes, temporal characteristics and more. We summarize the general approaches as follows.

\textbf{Feature based approaches} select features that have great impacts on information propagation, such as the content of information \cite{tsur2012s,ma2013predicting,hong2011predicting}, user characteristics \cite{cui2013cascading,bakshy2011everyone}, network structure \cite{romero2013interplay,weng2014predicting}, and temporal order \cite{pinto2013using}. These features are used as inputs of the machine learning model to obtain the final prediction results. Therefore, selecting the right feature or the right combination of features is crucial for the outcome. \cite{cheng2014can} finds that temporal and structural features are equally important in cascade prediction tasks. However, \cite{weng2014predicting} analyzes the impact of a comprehensive set of features on the popularity prediction and concludes that features based on community structure are the most powerful predictors. \cite{ shulman2016predictability} finds that the temporal feature has the most of the predictive power, that is, how quickly information reaches its first few adopters determines its influence. However, information usually contains a myriad of content, including images, audio, text, links and more, and each type of content has its own unique dissemination mechanism. A universal conclusion on the feature selection can hardly be made. Moreover, there are also multiple ways to quantify a feature. Therefore, while the feature-based approach has better interpretability, the handcrafted feature selection makes it hard to generalize.

\textbf{Model based approaches} assume that information is propagated according to a given pattern \cite{gomez2013modeling,ohsaka2017coarsening}, such as the independent cascade model \cite{goldenberg2001talk} and the linear threshold model \cite{kempe2003maximizing, ran2020generalized}. \cite{gomez2013modeling} uses survival theory to model the increase and decrease of a node's activation probability. \cite{ohsaka2017coarsening} used the vertex-weighted influence graph to approximate the diffusion properties of the input graph. \cite{lee2012modeling} borrows the idea from survival analysis to predict the likelihood that the content will be popular. \cite{shen2014modeling} first uses the probability model of reinforced Poisson processes to model the change of information popularity, and directly simulates the arrival process of individual popularity. \cite{bao2015modeling} presents a popularity prediction model based on the self-excited Hawkes process, which distinguishes the incentive size of each forward and improves the performance. RC-Tweet \cite{lymperopoulos2021rc} predicts real-time popularity with minimal publicly available information by fitting the evolution patterns of tweets from the entire spectrum of popularity. Nevertheless, as the real propagation mechanism is anomalous \cite{foroozani2019anomalous}, it is difficult to fit the real diffusion by a definite predefined mode.

\textbf{Embedding based approaches} aim to embed features in the cascade into a high dimensional space. The vectorized feature can then be easily adopted in learning and prediction. To avoid technical difficulties in learning the network topology, most early attempts focus on the proximity of nodes (users). \cite{bourigault2014learning} models the information propagation as heat diffusion in a high dimensional latent space through which the node's representation is learned. Following this work, \cite{bourigault2016representation, liu2016learning, xie2021independent} make modifications to improve the performance. The introduction of the deep learning method makes the use of network topology, temporal order and other features much more convenient, giving rise to a quick shift in designing the model. Now, most works on this topic would rely on deep learning kernels.
\cite{li2017deepcas} obtains node embedding by DeepWalk \cite{perozzi2014deepwalk} and transforms the cascade as node sequences to learn the cascade representation. \cite{cao2017deephawkes} transforms the cascade into a set of diffusion paths, and uses GRU \cite{chung2014empirical}, sum pooling, and non-parametric time kernel to aggregate the contributions of early adopters. \cite{kefato2018cas2vec} uses the recurrent neural network to learn the embedding solely from the cascade time series and predict whether a cascade is going to viral or not.

As the information spreading is controlled by both the underlying mechanism and network, recent progress starts to combine both the temporal (the dynamics) and structure (the network) feature in a cascade. \cite{chen2019information} samples a cascade graph as a series of sequential subcascade and adopts a dynamic multi-directional GCN to learn structural information of cascades. \cite{cao2020popularity} applies specifically designed graph neural network models to capture the change of node state as well as the network structure. \cite{zhao2020deep} uses structural property and the order of cascaded nodes to predict the future sequence of cascades. \cite{xu2020casgcn} combines the representation of network and the time of retweet to predict the future spreading size. \cite{wang2021ccasgnn} introduces a collaborative framework of GAT and GCN and stacks positional encoding into the layers of graph neural networks to improve the performance. \cite{feng2021aecasn} learns the structural characteristics and dynamic features of whole cascade networks, maps the cascades into low-dimensional vectors and then predicts the cascade size.

While the embedding of the network structure and the cascade temporal sequence can be readily obtained independently, how to best learn and combine these two interdependent properties remains to be explored. In this paper, we aim to make improvements in this direction. As we show later, by properly constructing the input and the learning kernel, the performance can be significantly improved without extra computational cost.

\section{Model Overview}
\subsection{Problem Definition}
Because the whole social network is too big to learn directly, we often use the cascade graph instead which contains sufficient topological information for prediction.

\textbf{Definition 1 Cascade graph} Let $G=\left( V,E \right)$ be a static social network, where $V$ is the set of all nodes (users) and $E$ is the set of all edges. One message will reach a certain part of the network, giving rise to a subset of nodes that retweet or adopt the message. Denote $V_i \subseteq V$ by the set of cascaded nodes for message $i$, the subgraph $C_i=\left( V_i,E_i \right) $ is defined as the cascade graph, where  $E_i \subseteq E$ is a set of all edges that connect between nodes in $V_i$. Because the spreading of a message is time dependent, we denote $C_i^T=\left( V_i^T,E_i^T \right) $ by a cascade graph of message $i$ given the observation time window $T$ from the beginning of the spreading.

\textbf{Definition 2 Activation state} We consider a node is activated (state 1) when it retweets the message, which can be equivalently interpreted as that the message reaches this node. A node is inactive (state 0) when it is not activated yet. The state vector $B_{i}^{T}(t)=\left\{ \left( {{b}_{1}},{{b}_{2}},\ldots ,{{b}_{|V_i^T|}} \right)\left| {{b}_{l}}=0\text{ }or\text{ }1,\text{ }l\in (1, 2 \ldots ,|V_i^T|) \right. \right\}$ captures the state of nodes at time $t$ in a cascade graph $C_i^T$.

\textbf{Definition 3 Cascade snapshot} By combining the observed cascade graph $C_i^T$ and the state vector at any given time $t$ during the observation, we can build a snapshot of the cascade graph ${S_i}^T(t)=\left\{ V_i^T,E_i^T,B_i^{T}(t) \right\}$. It captures all topological information in the observation and the nodes' state information at a particular time.

\begin{figure}[h]
	\centering
	\includegraphics[width=\linewidth]{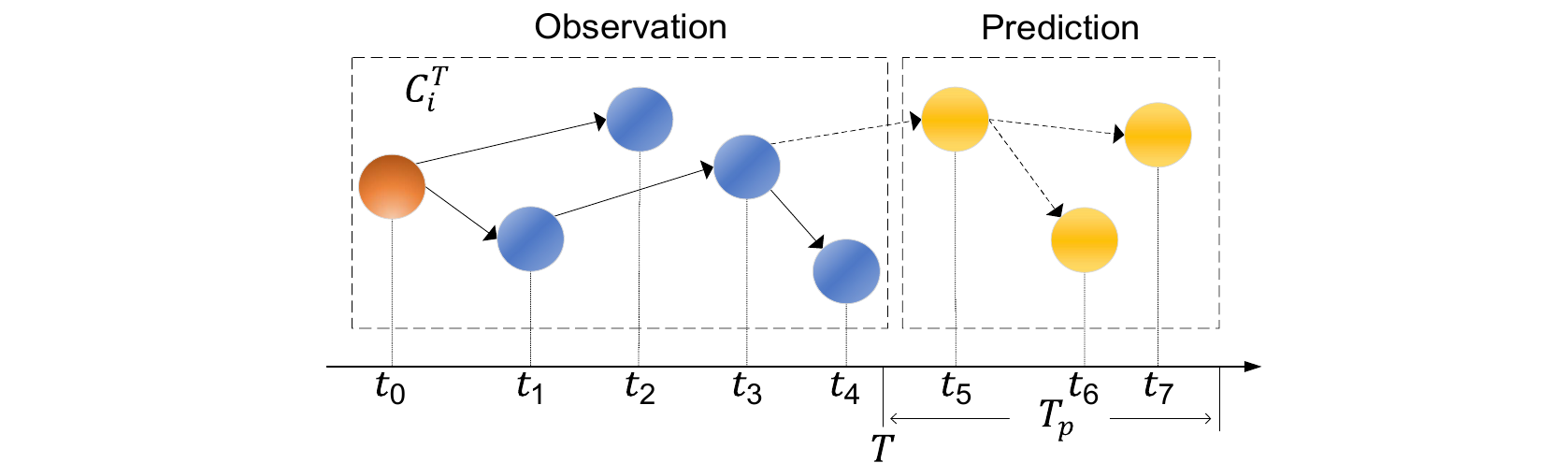}
	\caption{An illustration of the information cascade.}
	\label{fig-cascades-graph}
	%	\Description{information diffusion process.}
\end{figure}

In this study, we are interested in the size of the cascade defined as the number of retweets or equivalently the number of nodes (users) reached by a message. More specifically, we predict the growth size $\Delta R_{i}^{{{T}_{p}}}=\left| R_{i}^{T+{{T}_{p}}} \right|-\left| R_{i}^{T} \right|$ of a cascade after an observation period $T$, as shown in Figure \ref{fig-cascades-graph}. The same task is also investigated in other previous studies \cite{li2017deepcas,cao2017deephawkes,chen2019information}. 

\subsection{Model Design}

\begin{figure*}
	\includegraphics[width=\textwidth]{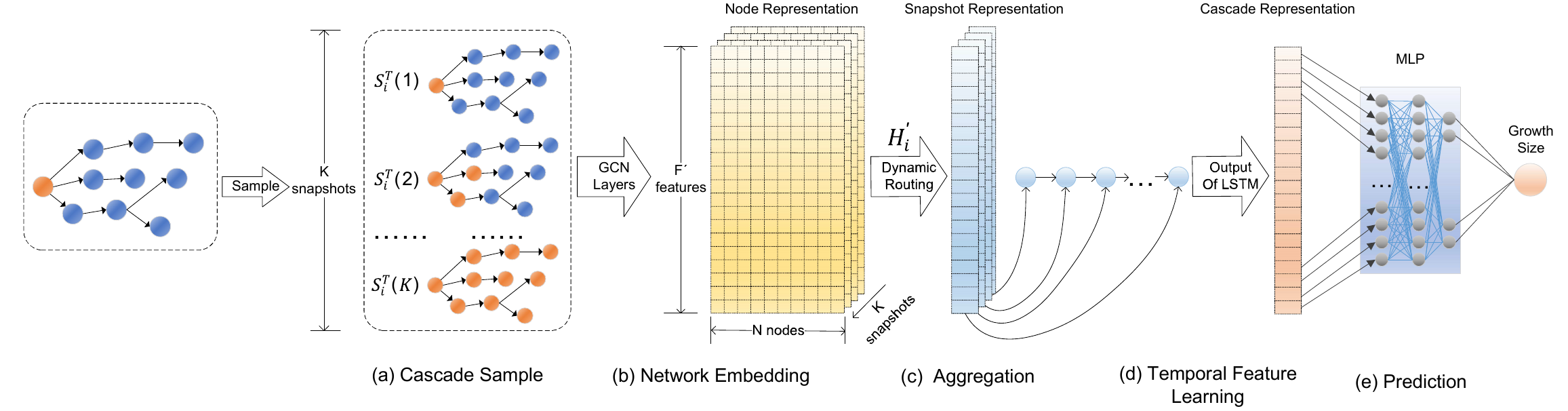}
	\caption{ Framework of CasSeqGCN.}
	\label{fig-framework}
\end{figure*}

The input of CasSeqGCN is an observed cascade sequence up to time $T$ and the corresponding cascade graph. The output is the cascade growth size $\Delta R_{i}^{T_p}$. As illustrated in Figure \ref{fig-framework}, CasSeqGCN contains five parts. (a) Cascade Sample: we obtain a sequence of cascade snapshots from the input. (b) Network Embedding: GCN is utilized to embed each cascade snapshot. (c) Aggregation: a dynamic routing method is applied to calculate the weight of each node, which is used to combine node vectors of one snapshot into one vector. (d) Temporal Features Learning: the sequence of snapshot vectors is fed into the LSTM layer to learn the temporal feature. (e) Prediction: a Multi-Layer Perceptron (MLP) is used to obtain the prediction result of cascade growth size.

\subsubsection{Cascade Sample}
\begin{figure}
	\centering
	\includegraphics[width=\linewidth]{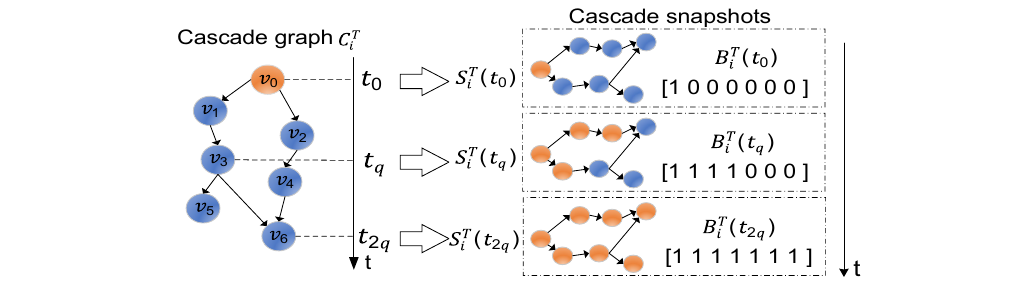}
	\caption{Cascade sample.}
	\label{fig-cascade-sample}
\end{figure}

Using the observed cascade graph $C_i^T$ and the cascade sequence, we can generate a sequence of cascade snapshot $\{ S_i^T(t) \}$. Each cascade snapshot has the same network topology, but the state vector may vary. The full sample strategy is to generate a new snapshot when a new node is activated, forming a snapshot sequence $\{ S_i^T(t_0), S_i^T(t_1), S_i^T(t_2), \dots \}$ with size $|V_i^T|$. To save the computation cost, we apply a partial sample strategy in this study in which the snapshot is taken from the first to the last cascade with increment $q$, forming a snapshot sequence $\{ S_i^T(t_0), S_i^T(t_q), S_i^T(t_{2q}), \dots \}$. Figure \ref{fig-cascade-sample} illustrates an example when $q=3$. Totally, there will be $K$ snapshots where,
\begin{equation*}
	K=1+ \lceil{ \frac{|V_i^T| -1}{q}}\rceil.
\end{equation*}

\subsubsection{Network Embedding}
In order to capture the structural characteristics of the cascade graph and the dynamical change of node state, we choose GCN among the existing state-of-the-art GNN models. GCN is based on neighbor aggregation by defining a spectral operation in Fourier domain for the convolutional operation on graphs. To briefly summarize, GCN studies the properties of graphs with the help of the eigenvalues and eigenvectors of graph Laplace matrices. For a cascade snapshot ${{S}_{i}}^T\left(t\right)=\left( {{V}_{i}}^T,{{E}_{i}}^T,B_{i}^{T}\left(t\right) \right)$, the input to the GCN layer consists of two parts: a vertex feature matrix $H\in {{\mathbb{R}}^{n\times F}}$ and an adjacency matrix $A\in {{\mathbb{R}}^{n\times n}}$ of the cascade graph, where $n$ is the number of vertices, $F$ is the number of features. Each row of $H$ is associated with a vertex feature, such as its activation state, degree and so on. More specifically, the GCN layer outputs ${H}'\in {{\mathbb{R}}^{n\times {F}'}}$ through the operation,
\begin{equation}
	{H}'=\sigma \left( {{L}^{sn}}H{{W}^{\intercal }}+b \right).
\end{equation}
${{L}^{sn}}$ is the symmetric normalized Laplace matrix defined as:
\begin{equation}
	{{L}^{sn}}={{D}^{-1/2}}L{{D}^{-1/2}}=\text{I}-{{D}^{-1/2}}A{{D}^{-1/2}},
\end{equation}
where $L$ is the Laplace matrix, $D$ is the degree matrix (diagonal matrix) of the vertices and $I$ is the identity matrix. Since GCN can only deal with undirected graphs, we include the node's in-degree and out-degree in the feature matrix $H$ to reflect the direction of influence.

\subsubsection{Aggregation}
To make use of the embedded vector of $N = |V_i^T|$ nodes, we aggregate $N$ vectors into one as the representation of the cascade snapshot. Here, we design a novel aggregation method named dynamic routing to calculate the weight of each node, which is inspired by the dynamic routing algorithm used in the capsule network \cite{sabour2017dynamic}. The weight coefficient reflects the contribution of the node in the graph representation. If the node representation is more similar to the graph representation, it will get a larger weight coefficient. The effect is amplified in each iteration, and the final coefficient is obtained through $r$ iterations. In particular, we first perform a linear affine transformation on the node representation vectors:
\begin{equation}
	U=W{H}',
\end{equation}
where $W$ is the mapping matrix, ${H}'$ is the node representation matrix. The output of each dynamic routing is:
\begin{equation}
	{{v}_{j}}=\underset{i}{\mathop \sum }\,{{c}_{ij}}{{u}_{i}},
\end{equation}
where ${{c}_{ij}}$ is the weight coefficient of node $i$ in the $j^\text{th}$ iteration of dynamic routing, ${{u}_{i}}$ is the representation vector after affine transformation of user $i$. The calculation of ${{c}_{ij}}$ is:
\begin{equation}
	{{c}_{ij}}=\frac{\text{exp}\left( {{b}_{ij}} \right)}{\mathop{\sum }_{k}\text{exp}\left( {{b}_{ik}} \right)},{{b}_{ij}}=cos\_sim\left( {{u}_{i}},{{v}_{j-1}} \right),
\end{equation}
where $cos\_sim$ is the calculation of cosine similarity between vectors. All ${{b}_{ij}}$'s are initialized to 0 at first.

\begin{algorithm}
	\caption{Dynamic routing algorithm}
	\begin{algorithmic}[1]
		\Require Iteration number: $r$, User embedding matrix: ${H}'$, User number: $N$
		\Ensure Snapshot representation vector: ${{v}_{j}}$
		\State Linear affine transformation: $U=W{H}'$
		\State for $i$ in $N$: ${{b}_{i0}=0}$
		\For{$j=1$ to $r$}
		\State node weight: ${{c}_{ij}} \gets softmax\left( {{b}_{ij}} \right)$
		\State snapshot representation: ${{v}_{j}} \gets \sum\nolimits_{i}{{{c}_{ij}}{{u}_{i}}}$
		\State ${{b}_{ij}}$ update: ${{b}_{ij}} \gets cos\_sim\left( {{u}_{i}},{{v}_{j-1}} \right)$
		\EndFor \\
		\Return ${{v}_{j}}$
	\end{algorithmic}
\end{algorithm}

\subsubsection{Temporal Feature Learning}
The temporal feature play a crucial role in the prediction of information diffusion \cite{cheng2014can}. For example, the time interval of retweets at a message's early stage is found to be a good predictor for its overall extent of spreading \cite{shulman2016predictability}. In our model, the temporal information is preserved in the sequence of the embedded cascade snapshots. To make use of such information, we apply the Long Short-Term Memory (LSTM) kernel in this part. We calculate the four states as follows:
\begin{equation}
	\begin{matrix}
		z=\tanh \left( W\left( {{v}^{t}}\parallel {{h}^{t-1}} \right) \right),  \\
		{{z}^{m}}=\sigma \left( {{W}^{m}}\left( {{v}^{t}}\parallel {{h}^{t-1}} \right) \right),  \\
		{{z}^{f}}=\sigma \left( {{W}^{f}}\left( {{v}^{t}}\parallel {{h}^{t-1}} \right) \right),  \\
		{{z}^{o}}=\sigma \left( {{W}^{o}}\left( {{v}^{t}}\parallel {{h}^{t-1}} \right) \right),  \\
	\end{matrix}
\end{equation}
where $\parallel$ denotes the vector concatenation operation, ${{v}^{t}}$ is the input of current unit, ${{h}^{t-1}}$ is the output of previous unit, ${{z}^{f}}$, ${{z}^{m}}$ and ${{z}^{o}}$ are the gate values between 0 and 1, $\sigma$ is the sigmoid activation function.
There are three main stages in time series feature extraction. In the forgetting stage, $z^f$ controls the output of previous unit $c^{t-1}$. The current input $v^t$ is selected by $z^m$ at the memory stage. We add the results of these two stages to get the $c^t$,
\begin{equation}
	{{c}^{t}}={{z}^{f}}\odot {{c}^{t-1}}+{{z}^{m}}\odot z,
\end{equation}
where $\odot $ is the entry-wise product. In the output stage, the result is mainly controlled by $c^t$ and $z^o$,
\begin{equation}
	\begin{matrix}
		{{h}^{t}}={{z}^{o}}\odot \tanh \left( {{c}^{t}} \right),  \\
		{{y}^{t}}=\sigma \left( {W}'{{h}^{t}} \right).  \\
	\end{matrix}
\end{equation}

\subsubsection{Prediction}
The output ${{y}^{t}}$ from the LSTM layer is fed into the Multi-Layer Perceptron (MLP) layer to get the final prediction as:
\begin{equation}
	\Delta \hat{R}_{i}^{T_p}=MLP\left( y_{i}^{t} \right).
\end{equation}
The loss function to be minimized is defined as:
\begin{equation}
	L\left( \Delta R_{i}^{T_p},\Delta \hat{R}_{i}^{T_p} \right)=\frac{1}{N}\underset{i=1}{\overset{N}{\mathop \sum }}\,{{\left( {{\log }_{2}}\Delta \hat{R}_{i}^{T_p}-{{\log }_{2}}\Delta R_{i}^{T_p} \right)}^{2}},
\end{equation}
where $N$ is the total number of cascades, $\Delta R_{i}^{T_p}$ and $\Delta \hat{R}_{i}^{T_p}$ are the true and predicted growth size for cascade ${{C}_{i}}^T$, respectively. Following \cite{li2017deepcas,cao2017deephawkes,chen2019information}, we use the log value of the growth size.

\section{Experiment}

\begin{table*}
	\centering
	\caption{ Statistics of the datasets.}
	\begin{tabular}{c|ccc|ccc|ccc}
		\toprule
		\multicolumn{1}{c}{DataSet} & \multicolumn{3}{|c|}{Weibo} & \multicolumn{3}{c|}{DBLP} & \multicolumn{3}{c}{Synthetic} \\
		\midrule
		All-nodes & \multicolumn{3}{c|}{1,776,950} & \multicolumn{3}{c|}{3,272,991} & \multicolumn{3}{c}{880} \\
		\midrule
		All-edges & \multicolumn{3}{c|}{308,489,739} & \multicolumn{3}{c|}{8,466,859} & \multicolumn{3}{c}{1992} \\
		\midrule
		${{T}_{p}}$     & \multicolumn{1}{c|}{9 hours} & \multicolumn{1}{c|}{12 hours} & \multicolumn{1}{c|}{24 hours} & \multicolumn{1}{c|}{1 year} & \multicolumn{1}{c|}{2 years} & \multicolumn{1}{c|}{3 years} & \multicolumn{1}{c|}{1 step} & \multicolumn{1}{c|}{2 steps} & \multicolumn{1}{c}{3 steps} \\
		\midrule
		Number of cascades & 29,123 & 29,122 & 34,897 & 30,106 & 29,998 & 29,991 & 13,024 &  12,204 & 10,584 \\
		Avg. observed nodes & 39.005 & 38.018 & 26.977 & 32.088 & 31.665 & 31.226 & 37.514 & 37.016 & 34.206 \\
		Avg. observed edges & 36.254 & 37.323 & 37.444 & 60.009 & 58.556 & 57.013 & 43.544 & 51.412 & 57.709 \\
		Avg. growth size & 4.874 & 6.999 & 20.616 & 1.965 & 2.101 & 8.578 & 1.900 & 6.460 & 13.596 \\
		\bottomrule
	\end{tabular}%
	\label{tab-datasets}%
\end{table*}%

\subsection{Datasets}
The performance of CasSeqGCN is evaluated in two real-world datasets and one synthetic dataset \footnote{\url{https://aminer.org}}. The two real-world datasets that are commonly utilized in related studies providing a nice data baseline for performance comparison. The synthetic data, in which the network structure and the spreading mechanism are explicitly given, are used to eliminate the potential bias due to the network structure loss and the noise in the spreading pattern. The statistics of the datasets as given in Table \ref{tab-datasets}.

\begin{figure}[htbp]
	\centering
	\subfigure[Weibo ($T_p = 12$).]{
		\begin{minipage}[t]{0.32\linewidth}
			\centering
			\includegraphics[width=0.9\linewidth]{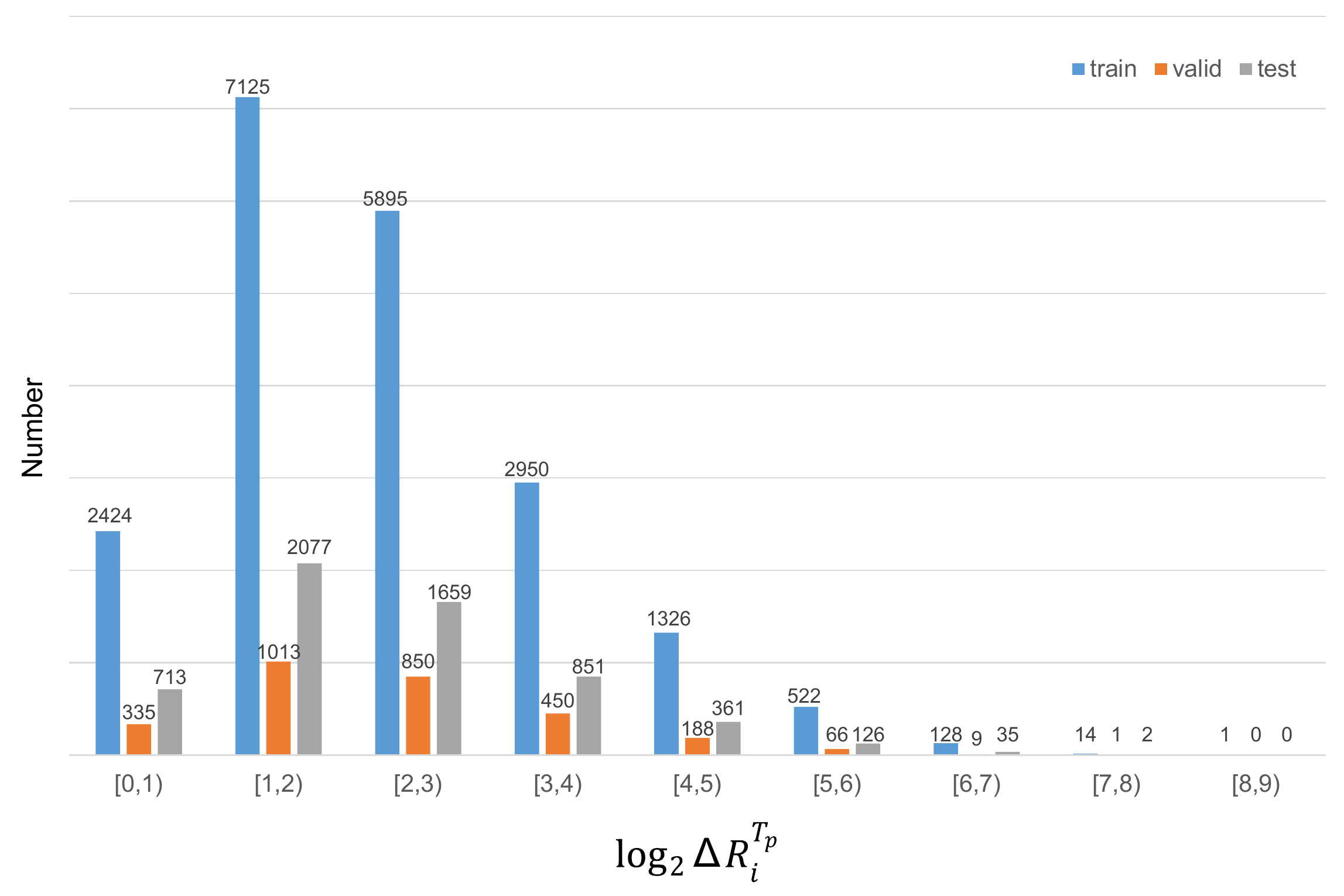}
			%				\caption{fig1}
		\end{minipage}%
	}
	\subfigure[DBLP ($T_p = 2$).]{
		\begin{minipage}[t]{0.32\linewidth}
			\centering
			\includegraphics[width=0.9\linewidth]{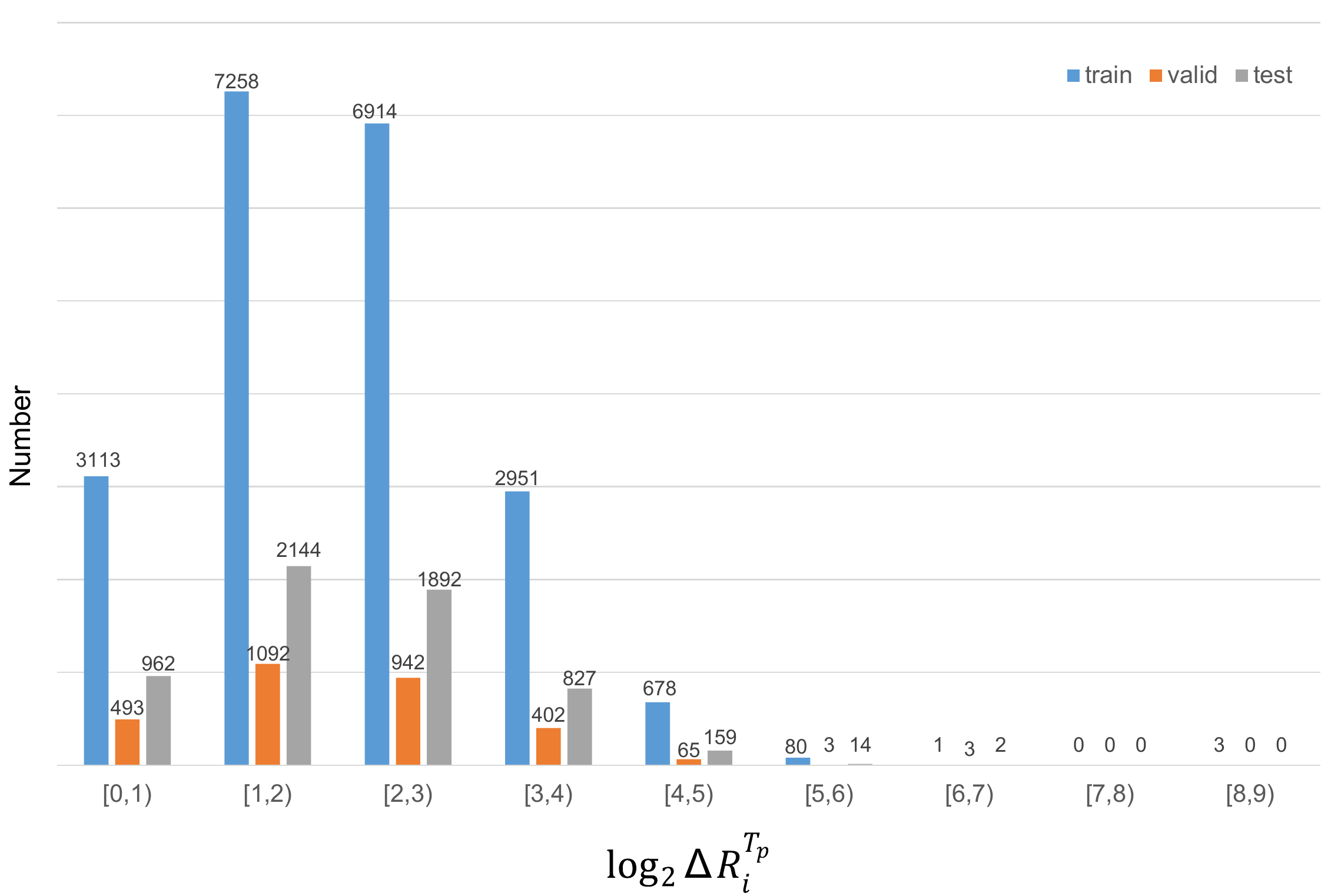}
			%				\caption{fig2}
		\end{minipage}%
	}
	%这个回车键很重要 \quad也可以
	\subfigure[Synthetic ($T_p = 2$).]{
		\begin{minipage}[t]{0.32\linewidth}
			\centering
			\includegraphics[width=0.9\linewidth]{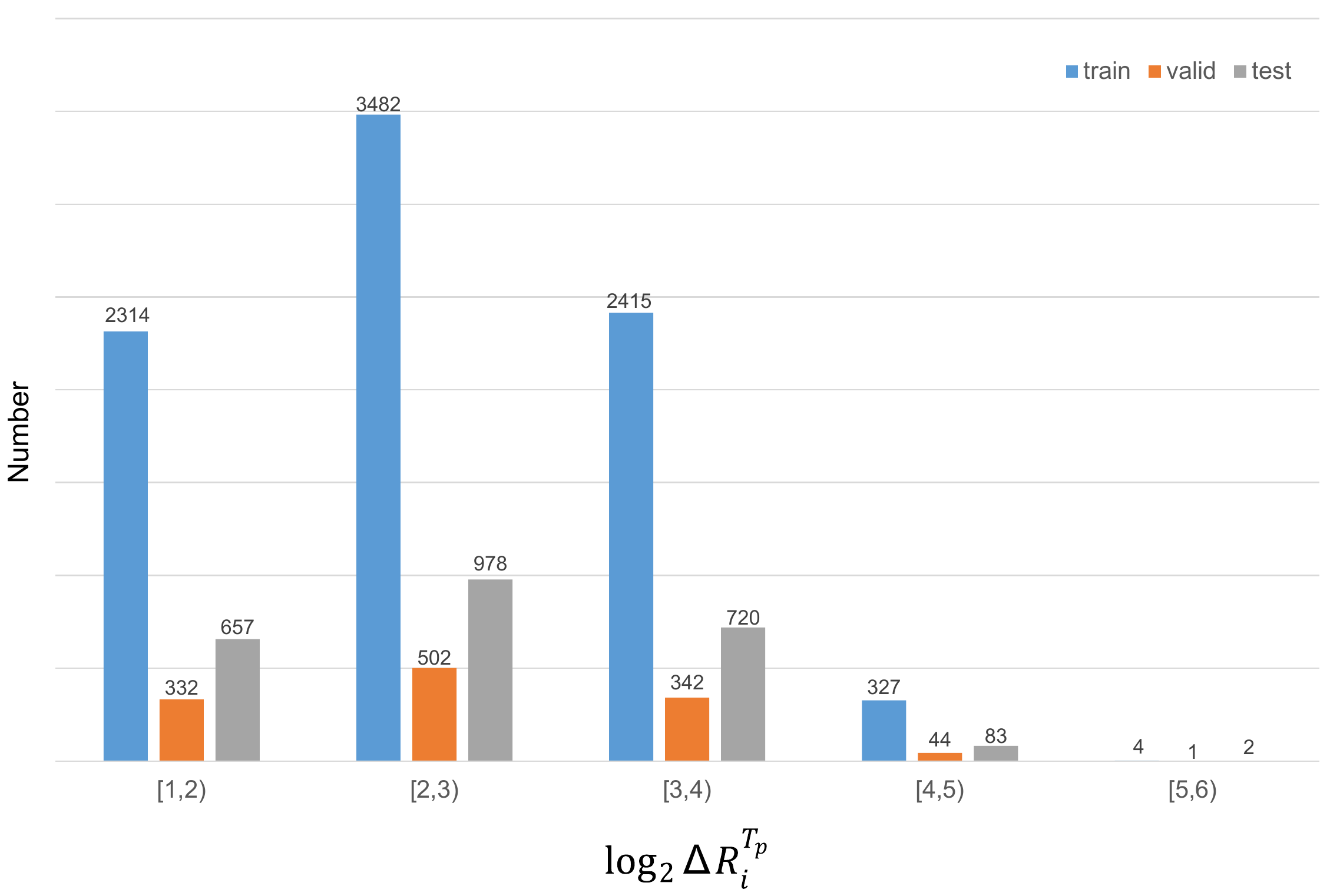}
			%				\caption{fig2}
		\end{minipage}
	}%
	\centering
	\caption{The distribution of growth size.}
	\label{fig-growth-size}
\end{figure}

Sina Weibo is the most popular Chinese microblogging platform with over 200 million daily active users. The network in this dataset is the following network, reflecting the relationship between the follower and the followee. If user A follows user B, there is a directed link from B to A. The network is composed of 1.78 million nodes and 308 million links. The data also contains spreading trajectories of over 300 thousand messages. We filter out cascades with fewer than 10 retweets in our experiment. We also consider the spreading is ended if a retweet does not occur for 12 hours. We predict the growth size in the last 9 hours, 12 hours and 24 hours respectively and randomly select 70\% for training, 10\% for validation and the remaining 20\% for testing.

Although the dynamics underlying the citation network are very different from information spreading, the assumption is similar that the network structure and the temporal order at the early stage are associated with a paper's future number of citations. Therefore, the citation data are also used in previous studies \cite{li2017deepcas,chen2019information}. When paper B appears in the reference list of paper A, there is a directed link from B to A. The DBLP citation data contains over 3 million nodes and over 8 million links. The cascade sequence here corresponds to a source paper and the subsequent papers that cite the source paper. The cascade graph corresponds to the co-citation relationship among papers in the cascade sequence. We filter out cascade sequences with the length fewer than 10 in our experiment. We also truncate the sequence if a paper does not get any new citation for 3 years. We predict the growth size in the last 1 year, 2 years and 3 years respectively and randomly select 70\% for training, 10\% for validation and the remaining 20\% for testing.

We generate a scale-free network by Barabasi–Albert model \cite{barabasi1999emergence}. The largest connected component contains 880 nodes and 1992 links. We use the independent cascade (IC) model \cite{goldenberg2001talk} and linear threshold (LT) model \cite{kempe2003maximizing, ran2020generalized} to model the information spreading starting at a randomly chosen seed node. We assign each link a random weight $w\in \left( 0,1 \right)$. For the LT model, we also assign each node a random threshold $\Gamma\in \left( 0,1 \right)$ corresponding to the minimum sum of activated link wights to activate this node and its links. The simulation is controlled by discrete time steps. The cascades with fewer than 3 time steps and 10 nodes are filtered out. We predict the growth size in the last 1 step, 2 steps and 3 steps, respectively.

The prediction tasks can be generally divided into two categories, depending on the choice of the observation time window $T$ and the prediction time $T_p$. One is to fix $T_p$ and let $T$ varies for each individual spreading trajectory \cite{zhou2020continual}. The other is to fix both $T$ and $T_p$ \cite{chen2019information}. We report the former in the main text and the latter in the Appendix. In particular, the observation window $T$ for each sequence is set by subtracting $T_p$ from the time of the last participant (Figure \ref{fig-cascades-graph}). For Weibo and DBLP data, the cascade snapshot is taken with a fixed increment $q$. But in the traditional IC and LT model, the order of nodes in the same time step can not be explicitly determined \cite{ran2020generalized}. Therefore, we use the time step to take the cascade snapshot in synthetic data, i.e. a new snapshot is taken at the end of one time step.

\subsection{Baselines}
\textbf{Feature based} method relies on pre-selected features to make the prediction. Following \cite{cheng2014can,zhang2016structure,yano2010s}, we compose a feature vector with structure and temporal features. Structural features include the average in-degree and out-degree of the cascade graph, the number of nodes, the number of leaf nodes, and the number of edges. The average activation time of nodes is used as the temporal feature, which reflects the spreading speed. To calculate it, we first find the time takes from the initial node to the $n^\text{th}$ node in the cascade sequence, divide the time by $n$, and then find the average value from all nodes. The feature vectors are independently sent to two predictors \textbf{Feature-Linear} and \textbf{Feature-Deep}. Feature-Linear is a linear regression model with L2 regularization. Feature-Deep uses a two-layer fully connected neural network \cite{chen2019information}.

\textbf{DeepCas} \cite{li2017deepcas} is the first end-to-end model that applies deep learning technology to the cascade prediction problem. It generates a series of paths by random walk to learn the representation of the cascade graph. The embedding vectors are sent to a bi-directional GRU neural network model with an attention mechanism to get the prediction results.

\textbf{DeepHawkes} \cite{cao2017deephawkes} generates multiple node sequences based on the propagation cascade. After obtaining node embedding through a mapping matrix, the node vector is fed into GRU to get the sequence representation. The retweet contribution of these representation vectors is calculated by the Hawkes process that considers user influence, self-exciting and time decay. After a weighted sum pooling operation, the prediction result is acquired through neural networks.

\textbf{CasCN} \cite{chen2019information} is a graph convolutional network (GCN)-based model. The cascade graph is divided into multiple subgraphs according to the set of activated nodes at different times. Hence, the sequence of subgraphs contains both the structure and temporal information of the spreading. The representation of the subgraph is learned through a dynamic multi-directional graph convolution kernel, which is then sent to the LSTM layer to learn the temporal feature. Our model is similar to CasCN in the way that they both sample the cascade graph at different times to get the structure and temporal information, and they both use the LSTM layer for feature learning. The difference is that CasCN considers the time evolving network structure, while our model uses a fixed network structure given by the cascade graph $C_i^T$ in which the state of nodes varies with time. Indeed, because the network structure is changing in CasCN, the traditional GCN model can not be directly applied. The adjacency matrix changes at every time step, which makes the convolution operation a bit time consuming. At the same time, building the adjacency matrix for each subgraph also gives rise to greater space complexity.

\textbf{CoupledGNN} \cite{cao2020popularity} uses two coupled graph neural networks to capture the interplay between node activation states and the spread of influence. By stacking graph neural network layers, the method characterizes two critical components (states and influence) of cascading effect along with the network structure in a successive manner. CoupledGNN requires a feature matrix of each node. Given the size of Weibo and DBLP data, it is computationally expensive and very difficult to process such a large matrix. Therefore, we choose to test CoupledGNN in the synthetic dataset only.

\textbf{CasGCN} \cite{xu2020casgcn} uses the convolutional layer to learn the node embedding. The vector of each node is further merged with the activation time of the node to combine the structural and temporal information. While it shares a similar name with our method, the CasGCN and CasSeqGCN are different significantly. CasGCN learns only one network embedding and the temporal information is added as one extra dimension of the node representation. In contrast, CasSeqGCN learns multiple embeddings of the network from the spreading sequence. Moreover, CasGCN directly uses the aggregated representation vector for prediction, whereas CasSeqGCN applied the LSTM layer to specifically utilize the temporal information. Unfortunately, the code of CasGCN is not found publicly available, making it impossible to make a fair comparison. We try our best to follow the scheme of \cite{xu2020casgcn} and implement an approximate model that uses a GCN layer to learn the embedding and apply the attention mechanism to aggregate node representations. Given the fact that many details are not explicitly given in \cite{xu2020casgcn}, especially some subtle parts that may optimize the overall performance, the CasGCN results presented in this paper may be an approximation of what the original model could offer.

\textbf{CCasGNN} \cite{wang2021ccasgnn} learns the user embedding through a collaborative framework of GAT and GCN. It stacks positional encodings into the layers to ensure that positional information should be considered in every graph neural network layer. In addition, the multi-head attention mechanism is employed to capture the relationships between all nodes from a global perspective.

\textbf{AECasN} \cite{feng2021aecasn} divides the cascade into multiple time slides and obtains the preliminary representation using the out-degree of nodes at different time slides. The representation vector is multiplied by discrete variables of the time decay effect to approximate the temporal features and then fed into the autoencoder model for training and learning.

\subsection{Experimental Setup}
We choose MSLE as the evaluation metric which is commonly adopted for performance evaluation in the cascade prediction problem \cite{li2017deepcas,cao2017deephawkes,chen2019information}, which is also the loss function of the model. It is calculated as:
\begin{equation}
	MSLE=\frac{1}{N}\underset{i=1}{\overset{N}{\mathop \sum }}\,{{\left( {{\log }_{2}}\vartriangle \hat{R}_{i}^{T_p}-{{\log }_{2}}\vartriangle R_{i}^{T_p} \right)}^{2}},
\end{equation}
where $N$ is the total number of cascades, $\vartriangle R_{i}^{T_p}$ and $\vartriangle \hat{R}_{i}^{T_p}$ are the true and predicted growth size for message $i$, respectively.

The hyperparameters for the baseline models are selected as follows. For feature-based linear regression method, the L2-coefficient is {1, 0.5, 0.1, 0.01, 0.001, 0.0005}. For feature-based fully connected neural network, the hidden layer is set to 2, and the dropout value is chosen as {0.1, 0.3, 0.5, 0.7}. For DeepCas and DeepHawkes, the dimension of the node representation vector is set to 50. There are 32 units in the hidden layer of each GRU and the number of neurons in the two hidden layers of Multi-Layer Perceptron is 32 and 16 respectively. The learning rate of node embedding is $5\times {{10}^{-5}}$. The time interval of DeepHawkes is set to 3 hours in the Weibo data, 1 year in the DBLP data and 1 time step in the synthetic data. For CasCN, the number of GCN layers is set to 2, and other parameters are the same as those in DeepCas and DeepHawkes.

The detailed setup of our model is as follows. The learning rate is chosen from 0.005, 0.01, 0.03, 0.05. Both GCN and LSTM have 2 layers, each layer contains 32 units. The dimension of node and snapshot vectors is 32. The dropout rate of LSTM and MLP layers is 0.5 during training. The partial sample is done with increment $q=5$. The features taken in GCN layer include a node's in-degree, out-degree and its activation state. The dynamic routing is controlled by $r = 3$.

\subsection{Performance Comparison}

\begin{table*}[htbp]
	\centering
	\caption{Overall prediction performance.}
	\begin{tabular}{c|ccc|ccc|ccc}
		\toprule
		\multicolumn{1}{c|}{DataSet } & \multicolumn{3}{c|}{Weibo} & \multicolumn{3}{c|}{DBLP} & \multicolumn{3}{c}{Synthetic} \\
		\midrule
		Evaluation Metric & \multicolumn{9}{c}{MSLE} \\
		\midrule
		\diagbox{Models}{${{T}_{p}}$} & \multicolumn{1}{c|}{9 hours} & \multicolumn{1}{c|}{12 hours} & 24 hours & \multicolumn{1}{c|}{1 year} & \multicolumn{1}{c|}{2 years} & 3 years & \multicolumn{1}{c|}{1 step} & \multicolumn{1}{c|}{2 steps} & 3 steps \\
		\midrule
		Feature\_Linear & 1.045 & 1.196 & 1.724 & 0.366 & 0.814 & 0.887 & 0.316 & 0.545 & 0.631 \\
		Feature\_Deep & 0.981 & 1.186 & 1.636 & 0.310 & 0.666 & 0.866 & 0.299 & 0.533 & 0.615 \\
		DeepCas & 0.979 & 1.184 & 1.538 & 0.355 & 0.722 & 0.874 & 0.289 & 0.493 & 0.517 \\
		DeepHawkes & 0.984 & 1.190 & 1.550 & 0.521 & 0.787 & 0.929 & 0.299 & 0.545 & 0.618 \\
		CasCN & 0.981 & 1.181 & 1.521 & 0.323 & 0.598 & 0.733 & 0.292 & 0.516 & 0.508 \\
		CoupledGNN & - & - & - & - & - & - & 0.867 & 0.953 & 1.128 \\
		CasGCN & 0.975 & 1.183 & 1.584 & 0.353 & 0.714 & 0.827 & 0.245 & 0.435 & 0.462 \\
		CCasGNN & 0.979 & 1.185 & 1.567 & 0.348 & 0.717 & 0.863 & 0.276 & 0.320 & 0.339 \\
		AECasN & 0.981 & 1.191 & 1.544 & 0.361 & 0.718 & 0.853 & 0.294 & 0.544 & 0.593 \\
		CasSeqGCN & \textbf{0.471} & \textbf{0.611} & \textbf{0.957} & \textbf{0.155} & \textbf{0.335} & \textbf{0.348} & \textbf{0.183} & \textbf{0.224} & \textbf{0.269} \\
		\bottomrule
	\end{tabular}%
	\label{tab-overall-performance}%
\end{table*}%

The performance comparison is shown in Table \ref{tab-overall-performance}. CasSeqGCN outperforms all baseline models in all three datasets,  improving about 40\% over the best baseline method on the Weibo and DBLP, and has also achieved great results on the synthetic data. The MSLE for some methods is smaller than those reported in \cite{li2017deepcas} with similar dataset. This is probably because the observation time window is not fixed, which makes the prediction task easier compared with that in previous studies \cite{li2017deepcas,cao2017deephawkes,chen2019information}.

In general, the deep learning method demonstrates a clear advantage. For the same input, Feature\_Deep is always better than Feature\_Linear. It is also interesting to note that the feature based method, despite its simplicity, can sometimes be as good as or better than more sophisticated methods. In DeepCas, the node representation is learned by Node2Vec \cite{grover2016node2vec}. As the node sequences extracted through the random walk will not be updated continuously with the training of the model, the performance of the prediction is limited. DeepHawkes introduces the Hawkes process on the basis of deep learning technology. But the assumption of the Hawkes process is strong and the real spreading mechanism can be more complicated and random. Consequently, the performance of DeepHawkes is sometimes worse than feature based ones.
CasCN learns the changing topology of the cascade graph at different observation points. It introduces a very nice approach to convert an adjacency matrix to a vector. However, it seems that the evolution of network structure is not the best feature for prediction. Indeed, although the cascaded nodes change with time, the underlying network does not. Therefore, using a fixed network structure and evolving node state would be a better combination, as applied in CasSeqGCN. CoupledGNN leverages two specifically designed graph neural networks to capture the cascading effect. The underlying network topology, however, is not fully utilized in this model, making the performance not as good in the small synthetic dataset. The computational cost also makes it hard to implement in large-scale data. The CasGCN combines the network embedding and the activation time of nodes. As the time is only one dimension within the high-dimensional vector of node representation, and there is no layer that specifically processes the activation time, the utilization of the temporal information is limited in CasGCN. Despite the similarity of their names, the prediction results by CasGCN are less accurate than that by CasSeqGCN. CCasGNN uses a collaborative framework based on graph neural networks. However, the temporal information represented by positional encodings restricts the improvement of accuracy. Although AECasN takes into account the temporal order of nodes in a cascade, it does not fully take into account the network topology. In addition, the dynamic features are learned only by multiplying the weight coefficients. Hence, its performance is not as well as our model.

\subsection{Ablation Study}
The advanced performance of CasSeqGCN prompts us to ask, to what extent does each part of our model contribute to the final outcome. To answer this question, we perform ablation studies by comparing the CasSeqGCN with other variants.

We adopt a dynamic routing approach to aggregate vectors of nodes in the cascade graph. But there are many other approaches for vector aggregation. To exam the benefit of the dynamic routing approach, we consider the following alternative models.

\textbf{CasSeqGCN\_CN} uses the capsule network for aggregation. The capsule network method is widely applied in computer vision, in which the length and direction of output vectors are associated with the existence probability and posture information of features, respectively. Therefore, it is claimed to be able to well capture the presence of a feature. This can be very helpful in the cascade prediction task as it is believed that similar structural and temporal features would yield similar information popularity.

\textbf{CasSeqGCN\_MH} uses the multi-head attention mechanism \cite{vaswani2017attention} to calculate the attention coefficients between the pair of nodes for vector aggregation. The multi-head attention mechanism is based on multiple but independent runs of Scaled Dot-Product Attention operation, which effectively avoids the over-fitting issue.

\textbf{CasSeqGCN\_Mean} simply averages all vectors of node embedding to get the vector representation of a cascade snapshot. It represents a baseline when no dedicated treatment is carried out in the aggregation part.

\begin{table*}[htbp]
	\centering
	\caption{Prediction performance of variants.}
	\begin{tabular}{c|ccc|ccc|ccc}
		\toprule
		\multicolumn{1}{c|}{DataSet } & \multicolumn{3}{c|}{Weibo} & \multicolumn{3}{c|}{DBLP} & \multicolumn{3}{c}{Synthetic} \\
		\midrule
		Evaluation Metric & \multicolumn{9}{c}{MSLE} \\
		\midrule
		\diagbox{Models}{${{T}_{p}}$} & \multicolumn{1}{c|}{9 hours} & \multicolumn{1}{c|}{12 hours} & 24 hours & \multicolumn{1}{c|}{1 year} & \multicolumn{1}{c|}{2 years} & 3 years & \multicolumn{1}{c|}{1 step} & \multicolumn{1}{c|}{2 steps} & 3 steps \\
		\midrule
		CasSeqGCN & \textbf{0.471} & \textbf{0.611} & \textbf{0.957} & \textbf{0.155} & \textbf{0.335} & \textbf{0.348} & 0.183 & 0.224 & 0.269 \\
		CasSeqGCN\_CN & 0.491 & 0.636 & 0.970 & 0.160 & 0.345 & 0.355 & 0.208 & 0.244 & 0.278 \\
		CasSeqGCN\_MH & 0.483 & 0.615 & 0.962 & 0.162 & 0.350 & 0.356 & \textbf{0.136} & \textbf{0.202} & \textbf{0.222} \\
		CasSeqGCN\_Mean & 0.479 & 0.615 & 0.965  & 0.286 & 0.359  & 0.361 & 0.226 & 0.254 & 0.318 \\
		CasSeqGCN\_noLSTM & 0.542 & 0.687 & 1.001  & 0.196 & 0.365  & 0.382 & 0.235 & 0.323 & 0.353 \\
		\bottomrule
	\end{tabular}%
	\label{tab-variants-performance}%
\end{table*}%

The performance comparison presented in Table \ref{tab-variants-performance} indicates that the dynamic routing approach is more suitable in dealing with the vector aggregation in real data. It is interesting to note that the multi-head attention mechanism performs best in synthetic data, and CasSeqGCN is only the second best. This is likely due to the fact that classical IC and LT models tend to underestimate the temporal complexity in spreading. The rank of a set of nodes in the cascade sequence is likely to be fixed in a specific discrete time step \cite{ran2020generalized}. This strong temporal correlation would be further strengthened by the multi-head attention mechanism which takes the correlation between nodes into account. But the multi-head attention mechanism needs to calculate the attention coefficient between each pair of nodes with time complexity $\text{O}\left( {{N}^{2}} \right)$, the dynamic routing approach only optimizes the similarity between each node and the aggregated vector with time complexity $\text{O}\left( {{N}} \right)$. Therefore, the dynamic routing applied in CasSeqGCN is still an optimal approach for its overall performance and high time efficiency.
Finally, the performance of other variants over that of CasSeqGCN\_Mean clearly demonstrates the importance of embedding aggregation. But note that CasSeqGCN\_Mean, which technically does not take any special operation, still outperforms other baseline methods in Table \ref{tab-overall-performance}. This implies that the aggregation part is not the most important source of performance enhancement.

The temporal feature learning in our model is done by the LSTM kernel. To test the role of LSTM in our model, we compose another baseline model \textbf{CasSeqGCN\_noLSTM} in which the LSTM kernel is replaced by an average operation. As shown in Table \ref{tab-variants-performance}, the performance without LSTM kernel drops by 10\% or even more, demonstrating the important role LSTM plays in capturing the temporal evolution of the input. Indeed, the LSTM and other related variants have been intensively applied in tasks that require temporal feature learning \cite{tan2020syntactic,wang2017topological,li2017deepcas, cao2017deephawkes,chen2019information}, which self proves its high efficiency. Note that the MSLE by CasSeqGCN\_noLSTM is higher than that of CasSeqGCN\_Mean, indicating that CasSeqGCN benefits more from the temporal learning part than from the aggregation part.

\begin{figure}[h]
	\centering
	\includegraphics[width=\linewidth]{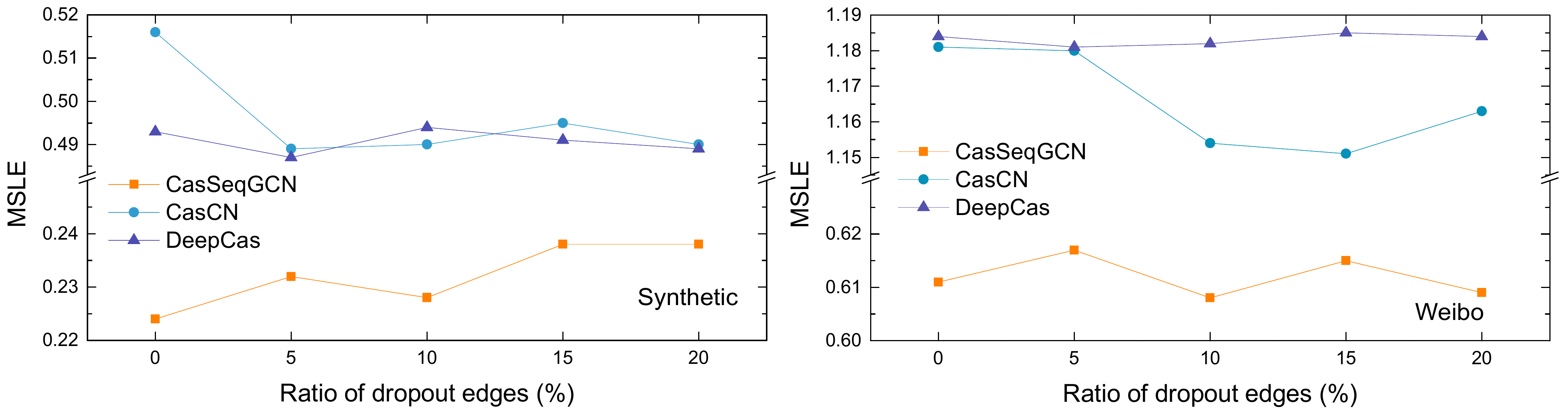}
	\caption{Impact of dropout edges on performance.}
	\label{fig-dropout-edges}
\end{figure}

A final question is, to what extent the model is affected by the data quality. We have shown that CasSeqGCN benefits from the GCN layer. However, because GCN relies on a node's neighboring nodes to learn its representation, missing edges between nodes may bring a big impact on the final outcome or even flip the performance rank in Table \ref{tab-overall-performance}. To answer this question, we randomly remove 5\%, 10\%, 15\%, and 20\% of edges in our data. For synthetic data, we direct remove edges from the network and predict the cascade growth size of the last 2 steps. For Weibo data, because the network is too sparse, we randomly remove edges from the cascade graph instead, and predict the cascade growth size of the last 12 hours. The results in Figure \ref{fig-dropout-edges} indicate that data with missing structure will bring fluctuations on the performance, but the improvement by CasSeqGCN does not shrink. CasSeqGCN still outperforms other methods.

\subsection{Discussions on Cascade Representation}
The above tests suggest that the design of the input and the use of GCN yield the most significant contribution to the overall performance improvement. Indeed, by focusing on the full cascade graph in observation, we preserve most structure information.

By introducing the activation state of nodes, the classical GCN can be applied and the temporal pattern is recorded. In this way, CasSeqGCN could learn a more accurate representation of the spreading than other baseline models. To compare the quality of the cascade representations learned by different approaches, we apply the representation learned by CasSeqGCN, DeepCas and CasCN to another type of task, which is to predict the information diffusion model in the synthetic data.

In particular, we generate synthetic spreading sequences using IC and LT models. In the training data, each sequence is labeled exactly by the generating model. The task is to predict the unknown label of a sequence in the test set. The cascade representation by CasSeqGCN, DeepCas and CasCN are sent to a two-layer fully connected neural network for classification. As the training data and the classifier are the same, the performance difference only reflects the different capabilities of the three models in learning the representation. As shown in Table \ref{model-prediction}, CasSeqGCN gives the best result, supporting our argument that CasSeqGCN has better performance because it learns the cascades better.

\begin{table}[htbp]
	\centering
	\caption{Diffusion model prediction on the synthetic dataset.}
	\footnotesize
	\resizebox{8cm}{1.5cm}{
		\begin{tabular}{l|rrr}
			\toprule
			Evaluation Metric & \multicolumn{3}{c}{AUC} \\
			\midrule
			\diagbox{Models}{${{T}_{p}}$} & \multicolumn{1}{l|}{1 step} & \multicolumn{1}{l|}{2 steps} & \multicolumn{1}{l}{3 steps} \\
			\midrule
			DeepCas & 0.702 & 0.758 & 0.781 \\
			CasCN & 0.591 & 0.652 & 0.683 \\
			CasSeqGCN & \textbf{0.781} & \textbf{0.872} & \textbf{0.851} \\
			\bottomrule
		\end{tabular}%
	}
	\label{model-prediction}%
\end{table}%

\subsection{Parameter Analysis}

\begin{figure}[h]
	\centering
	\includegraphics[width=\linewidth]{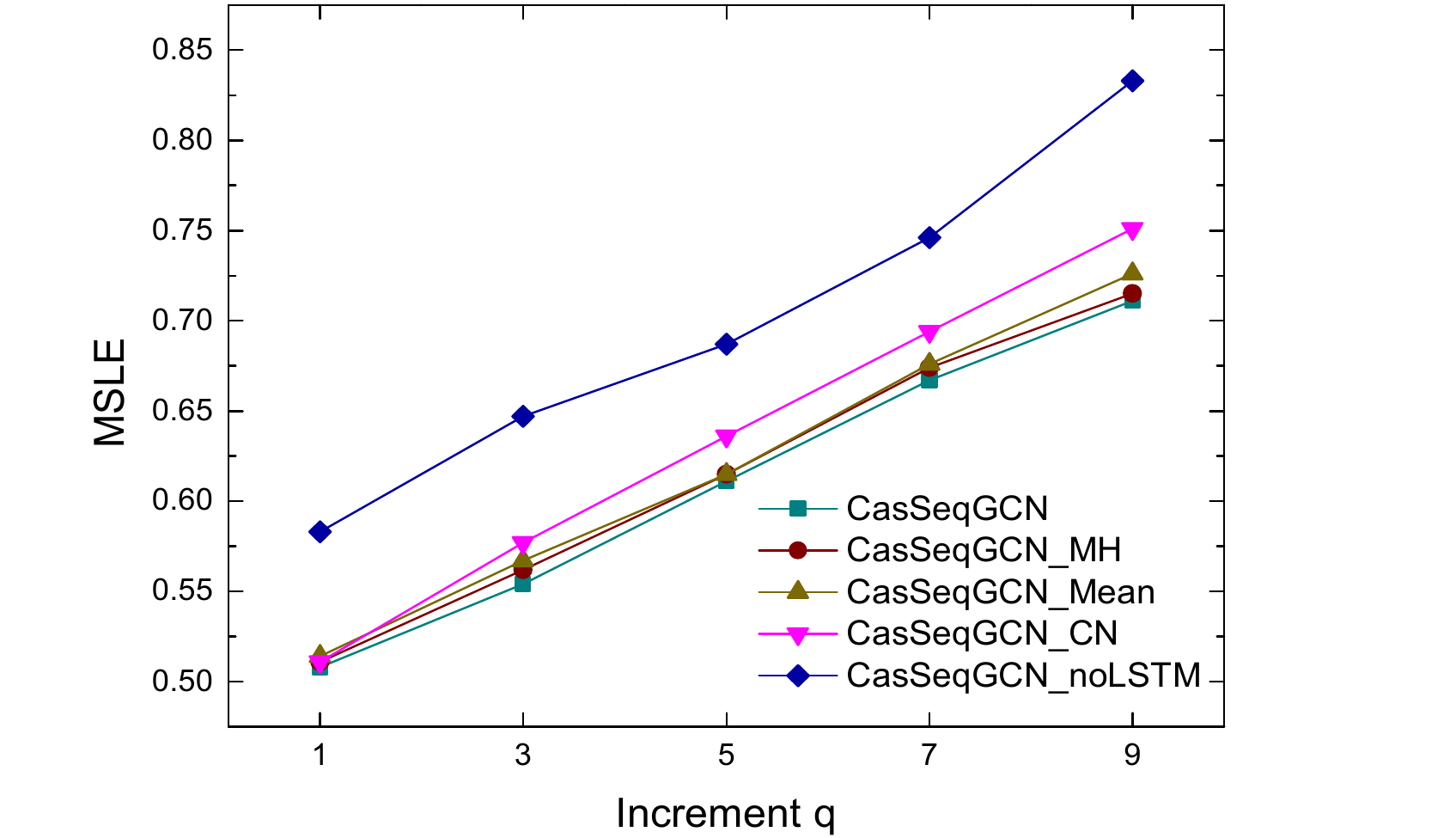}
	\caption{Impact of different increment $q$ on performance.}
	\label{fig-diffusion-increment}
\end{figure}

To save the cost of computation, we apply a partial sampling strategy by composing a subset snapshot sequence. The sampling increment is controlled by parameter $q$ which takes $q=5$ in this work. In Figure \ref{fig-diffusion-increment}, we plot the performance of CasSeqGCN and its variants with different $q$ values. These models are applied in Weibo to predict the cascade size in 12 hours. $q=1$ corresponds to the full sample, which gives the best performance but also the biggest computational cost. From $q = 1$ to $q = 5$, the MSLE increases by roughly 20\% from 0.508 to 0.611, but the computational cost is 5 times less. Therefore, one needs to choose $q$ value to balance the cost and the accuracy needed. The slope of different lines in Figure \ref{fig-diffusion-increment} is almost the same, indicating that the penalty is the same for different variants. Hence, different sample strategies only affect the quality of the embedded vectors, but do not change the aggregation and temporal feature learning part.

For the dynamic routing, we explore the impact of the number of routing $r$ on the prediction results. Results based on synthetic data and parameter $r=1, 2, 3, 4, 5$ are presented in Table \ref{tab-routing}. The optimal value is $r =3$, which is one taken in this study. Indeed, a small $r$ can bring the issue of under fitting that the coefficient is not optimized, while a large $r$ may assign too big a weight to a node that is close to the enter. Therefore, it is expected that the performance would peak at an intermediate $r$ value.

\begin{table}[htbp]
	\centering
	\caption{Impact of dynamic routing times $r$ on performance}
	\begin{tabular}{c|ccc}
		\toprule
		DataSet & \multicolumn{3}{c}{Synthetic} \\
		\midrule
		Evaluation Metric & \multicolumn{3}{c}{MSLE} \\
		\midrule
		\diagbox{$r$}{${{T}_{p}}$} & \multicolumn{1}{c|}{1 step} & \multicolumn{1}{c|}{2 steps} & 3steps \\
		\midrule
		1     & 0.206 & 0.252 & 0.304 \\
		2     & 0.193 & 0.247 & 0.292 \\
		3     & \textbf{0.183} & \textbf{0.227} & \textbf{0.269} \\
		4     & 0.199 & 0.231 & 0.286 \\
		5     & 0.207 & 0.227 & 0.288 \\
		\bottomrule
	\end{tabular}%
	\label{tab-routing}%
\end{table}%

\section{Conclusion and Future Work}
To summarize, we present CasSeqGCN, an end-to-end framework for cascade prediction. Using a fixed network structure and varying node states as the input, we utilize the classical GCN to learn the representation of each cascade snapshot. The representation accurately captures the structure information and preserves the temporal order of the spreading, allowing us to obtain an improved prediction result compared with several state-of-art methods. The method benefits from a novel approach to learn and predict information cascade, which is different from all existing ones and demonstrates good performance. Therefore, it can serve as a new baseline for future studies. Moreover, the context that the underlying network is given but the node activation states are varying does not only present in the problem of information spreading. In human mobility problem \cite{sun2013understanding, yan2017universal}, the transportation network is given and a traveler can be at different locations at different time. In studying the topic change of scientists \cite{jia2017quantifying, zeng2019increasing}, the knowledge graph is given and a scientist can switch from one area to another. A similar approach can be used to learn the representation of a traveler or a scientist. Therefore, our work brings more insights to a wider range of problems in the study of social system.

\appendix

\section{Cascade Prediction Using Fixed $T$ and $T_p$}

To avoid the forgetting problem mentioned in \cite{zhou2020continual}, we use a variable observation time window in the main text. But our model can also achieve good performance in the task based on the fixed observation time window $T$. We follow the same experimental setup in \cite{chen2019information}. 
%In the Weibo dataset, we set $T_p$ to be the next 24 hours and the length $T$ of the observation time window to be  $T =$ 1 hour, 2 hours, and 3 hours.
In the Weibo dataset, we extract all re-tweets of each post within the next 24 hours and set the length $T$ of the observation time window to be  $T =$ 1 hour, 2 hours, and 3 hours.
In the DBLP dataset, we track all citations of the paper within the next 10 years and set the observation time window $T$ to be 3, 4, and 5 years. In the synthetic dataset, the propagation process is controlled by discrete time steps, so we extract the cascades within the 6 time steps and set the observation time window $T$ to be 3, 4, and 5 steps. The results are shown in Table \ref{tab-addlabel}. Our approach outperforms other baseline methods, supporting its capability in cascade prediction.

\begin{table*}[htbp]
	\centering
	\caption{Overall prediction performance based on the fixed $T$ and $T_p$.}
	\begin{tabular}{c|ccc|ccc|ccc}
		\toprule
		\multicolumn{1}{c|}{DataSet } & \multicolumn{3}{c|}{Weibo} & \multicolumn{3}{c|}{DBLP} & \multicolumn{3}{c}{Synthetic} \\
		\midrule
		Evaluation Metric & \multicolumn{9}{c}{MSLE} \\
		\midrule
		\diagbox{Models}{$T$} & \multicolumn{1}{c|}{1 hour} & \multicolumn{1}{c|}{2 hours} & 3 hours & \multicolumn{1}{c|}{3 years} & \multicolumn{1}{c|}{5 years} & 7 years & \multicolumn{1}{c|}{3 steps} & \multicolumn{1}{c|}{4 steps} & 5 steps \\
		\midrule
		Feature\_Linear & 5.958 & 5.583 & 4.930 & 4.602 & 3.843 & 2.711 & 1.295 & 1.289 & 1.251 \\
		Feature\_Deep & 6.071 & 5.750 & 5.121 & 4.481 & 3.370 & 2.031 & 1.280 & 1.191 & 1.112 \\
		DeepCas & 4.790 & 4.456 & 3.761 & 3.082 & 2.233 & 1.573 & 0.970 & 0.858 & 0.747 \\
		DeepHawkes & 4.858 & 4.644 & 3.856 & 3.256 & 2.593 & 1.650 & 1.179 & 1.092 & 0.993 \\
		CasCN & 4.537 & 4.291 & 3.673 & 2.062 & 1.844 & 1.207 & 0.896 & 0.765 & 0.735 \\
		CoupledGNN & - & - & - & - & - & - & 0.961 & 0.847 & 0.741 \\
		CasGCN & 5.268 & 4.811 & 4.076 & 1.896 & 1.589 & 1.296 & 0.883 & 0.732 & 0.698 \\
		CCasGNN & 4.413 & 4.201 & 3.617 & 1.792 & 1.553 & 1.146 & 0.632 & 0.425 & 0.387 \\
		AECasN & 4.652 & 4.449 & 3.744 & 3.162 & 2.362 & 1.649 & 0.958 & 0.838 & 0.684 \\
		CasSeqGCN & \textbf{3.833} & \textbf{3.612} & \textbf{3.178} & \textbf{1.537} & \textbf{1.377} & \textbf{1.190} & \textbf{0.450} & \textbf{0.375} & \textbf{0.332} \\
		\bottomrule
	\end{tabular}%
	\label{tab-addlabel}
\end{table*}%

\bibliographystyle{unsrt}  


\begin{thebibliography}{1}

\bibitem{watts2016computational}
D.~Watts, Computational social science: Exciting progress and future
challenges, in: Proceedings of the 22nd ACM SIGKDD International Conference
on Knowledge Discovery and Data Mining, 2016, pp. 419--419.

\bibitem{zhang2016dynamics}
Z.-K. Zhang, C.~Liu, X.-X. Zhan, X.~Lu, C.-X. Zhang, Y.-C. Zhang, Dynamics of
information diffusion and its applications on complex networks, Physics
Reports 651 (2016) 1--34.

\bibitem{centola2010spread}
D.~Centola, The spread of behavior in an online social network experiment,
science 329~(5996) (2010) 1194--1197.

\bibitem{xie2021detecting}
J.~Xie, F.~Meng, J.~Sun, X.~Ma, G.~Yan, Y.~Hu, Detecting and modelling real
percolation and phase transitions of information on social media, Nature
Human Behaviour (2021) 1--8.

\bibitem{jia2015analysis}
T.~Jia, R.~F. Spivey, B.~Szymanski, G.~Korniss, An analysis of the matching
hypothesis in networks, PloS one 10~(6) (2015) e0129804.

\bibitem{kianian2021efficient}
S.~Kianian, M.~Rostamnia, An efficient path-based approach for influence
maximization in social networks, Expert Systems with Applications 167 (2021)
114168.

\bibitem{su2020emergence}
Z.~Su, C.~Gao, J.~Liu, T.~Jia, Z.~Wang, J.~Kurths, Emergence of nonlinear
crossover under epidemic dynamics in heterogeneous networks, Physical Review
E 102~(5) (2020) 052311.

\bibitem{cheng2014can}
J.~Cheng, L.~Adamic, P.~A. Dow, J.~M. Kleinberg, J.~Leskovec, Can cascades be
predicted?, in: Proceedings of the 23rd international conference on World
wide web, 2014, pp. 925--936.

\bibitem{zhou2021survey}
F.~Zhou, X.~Xu, G.~Trajcevski, K.~Zhang, A survey of information cascade
analysis: Models, predictions, and recent advances, ACM Computing Surveys
(CSUR) 54~(2) (2021) 1--36.

\bibitem{robles2020evolutionary}
J.~F. Robles, M.~Chica, O.~Cordon, Evolutionary multiobjective optimization to
target social network influentials in viral marketing, Expert Systems with
Applications 147 (2020) 113183.

\bibitem{wu2019dual}
Q.~Wu, Y.~Gao, X.~Gao, P.~Weng, G.~Chen, Dual sequential prediction models
linking sequential recommendation and information dissemination, in:
Proceedings of the 25th ACM SIGKDD International Conference on Knowledge
Discovery \& Data Mining, 2019, pp. 447--457.

\bibitem{zhao2020online}
L.~Zhao, J.~Chen, F.~Chen, F.~Jin, W.~Wang, C.-T. Lu, N.~Ramakrishnan, Online
flu epidemiological deep modeling on disease contact network, GeoInformatica
24~(2) (2020) 443--475.

\bibitem{tsur2012s}
O.~Tsur, A.~Rappoport, What's in a hashtag? content based prediction of the
spread of ideas in microblogging communities, in: Proceedings of the fifth
ACM international conference on Web search and data mining, 2012, pp.
643--652.

\bibitem{ma2013predicting}
Z.~Ma, A.~Sun, G.~Cong, On predicting the popularity of newly emerging hashtags
in t witter, Journal of the American Society for Information Science and
Technology 64~(7) (2013) 1399--1410.

\bibitem{cui2013cascading}
P.~Cui, S.~Jin, L.~Yu, F.~Wang, W.~Zhu, S.~Yang, Cascading outbreak prediction
in networks: a data-driven approach, in: Proceedings of the 19th ACM SIGKDD
international conference on Knowledge discovery and data mining, 2013, pp.
901--909.

\bibitem{weng2014predicting}
L.~Weng, F.~Menczer, Y.-Y. Ahn, Predicting successful memes using network and
community structure, in: Proceedings of the International AAAI Conference on
Web and Social Media, Vol.~8, 2014.

\bibitem{li2017deepcas}
C.~Li, J.~Ma, X.~Guo, Q.~Mei, Deepcas: An end-to-end predictor of information
cascades, in: Proceedings of the 26th international conference on World Wide
Web, 2017, pp. 577--586.

\bibitem{chen2019information}
X.~Chen, F.~Zhou, K.~Zhang, G.~Trajcevski, T.~Zhong, F.~Zhang, Information
diffusion prediction via recurrent cascades convolution, in: 2019 IEEE 35th
International Conference on Data Engineering (ICDE), IEEE, 2019, pp.
770--781.

\bibitem{zhao2019t}
L.~Zhao, Y.~Song, C.~Zhang, Y.~Liu, P.~Wang, T.~Lin, M.~Deng, H.~Li, T-gcn: A
temporal graph convolutional network for traffic prediction, IEEE
Transactions on Intelligent Transportation Systems 21~(9) (2019) 3848--3858.

\bibitem{bai2021a3t}
J.~Bai, J.~Zhu, Y.~Song, L.~Zhao, Z.~Hou, R.~Du, H.~Li, A3t-gcn: Attention
temporal graph convolutional network for traffic forecasting, ISPRS
International Journal of Geo-Information 10~(7) (2021) 485.

\bibitem{zheng2019addgraph}
L.~Zheng, Z.~Li, J.~Li, Z.~Li, J.~Gao, Addgraph: Anomaly detection in dynamic
graph using attention-based temporal gcn., in: IJCAI, 2019, pp. 4419--4425.

\bibitem{kipf2016semi}
T.~N. Kipf, M.~Welling, Semi-supervised classification with graph convolutional
networks, arXiv preprint arXiv:1609.02907.

\bibitem{hochreiter1997long}
S.~Hochreiter, J.~Schmidhuber, Long short-term memory, Neural computation 9~(8)
(1997) 1735--1780.

\bibitem{shulman2016predictability}
B.~Shulman, A.~Sharma, D.~Cosley, Predictability of popularity: Gaps between
prediction and understanding, in: Proceedings of the International AAAI
Conference on Web and Social Media, Vol.~10, 2016.

\bibitem{kefato2018cas2vec}
Z.~T. Kefato, N.~Sheikh, L.~Bahri, A.~Soliman, A.~Montresor, S.~Girdzijauskas,
Cas2vec: Network-agnostic cascade prediction in online social networks, in:
2018 Fifth International Conference on Social Networks Analysis, Management
and Security (SNAMS), IEEE, 2018, pp. 72--79.

\bibitem{liao2019popularity}
D.~Liao, J.~Xu, G.~Li, W.~Huang, W.~Liu, J.~Li, Popularity prediction on online
articles with deep fusion of temporal process and content features, in:
Proceedings of the AAAI Conference on Artificial Intelligence, Vol.~33, 2019,
pp. 200--207.

\bibitem{gou2018learning}
C.~Gou, H.~Shen, P.~Du, D.~Wu, Y.~Liu, X.~Cheng, Learning sequential features
for cascade outbreak prediction, Knowledge and Information Systems 57~(3)
(2018) 721--739.

\bibitem{hong2011predicting}
L.~Hong, O.~Dan, B.~D. Davison, Predicting popular messages in twitter, in:
Proceedings of the 20th international conference companion on World wide web,
2011, pp. 57--58.

\bibitem{kupavskii2013predicting}
A.~Kupavskii, A.~Umnov, G.~Gusev, P.~Serdyukov, Predicting the audience size of
a tweet, in: Proceedings of the International AAAI Conference on Web and
Social Media, Vol.~7, 2013.

\bibitem{cao2017deephawkes}
Q.~Cao, H.~Shen, K.~Cen, W.~Ouyang, X.~Cheng, Deephawkes: Bridging the gap
between prediction and understanding of information cascades, in: Proceedings
of the 2017 ACM on Conference on Information and Knowledge Management, 2017,
pp. 1149--1158.

\bibitem{bakshy2011everyone}
E.~Bakshy, J.~M. Hofman, W.~A. Mason, D.~J. Watts, Everyone's an influencer:
quantifying influence on twitter, in: Proceedings of the fourth ACM
international conference on Web search and data mining, 2011, pp. 65--74.

\bibitem{romero2013interplay}
D.~Romero, C.~Tan, J.~Ugander, On the interplay between social and topical
structure, in: Proceedings of the International AAAI Conference on Web and
Social Media, Vol.~7, 2013.

\bibitem{pinto2013using}
H.~Pinto, J.~M. Almeida, M.~A. Gon{\c{c}}alves, Using early view patterns to
predict the popularity of youtube videos, in: Proceedings of the sixth ACM
international conference on Web search and data mining, 2013, pp. 365--374.

\bibitem{gomez2013modeling}
M.~Gomez-Rodriguez, J.~Leskovec, B.~Sch{\"o}lkopf, Modeling information
propagation with survival theory, in: International conference on machine
learning, PMLR, 2013, pp. 666--674.

\bibitem{ohsaka2017coarsening}
N.~Ohsaka, T.~Sonobe, S.~Fujita, K.-i. Kawarabayashi, Coarsening massive
influence networks for scalable diffusion analysis, in: Proceedings of the
2017 ACM International Conference on Management of Data, 2017, pp. 635--650.

\bibitem{goldenberg2001talk}
J.~Goldenberg, B.~Libai, E.~Muller, Talk of the network: A complex systems look
at the underlying process of word-of-mouth, Marketing letters 12~(3) (2001)
211--223.

\bibitem{kempe2003maximizing}
D.~Kempe, J.~Kleinberg, {\'E}.~Tardos, Maximizing the spread of influence
through a social network, in: Proceedings of the ninth ACM SIGKDD
international conference on Knowledge discovery and data mining, 2003, pp.
137--146.

\bibitem{ran2020generalized}
Y.~Ran, X.~Deng, X.~Wang, T.~Jia, A generalized linear threshold model for an
improved description of the spreading dynamics, Chaos: An Interdisciplinary
Journal of Nonlinear Science 30~(8) (2020) 083127.

\bibitem{lee2012modeling}
J.~G. Lee, S.~Moon, K.~Salamatian, Modeling and predicting the popularity of
online contents with cox proportional hazard regression model, Neurocomputing
76~(1) (2012) 134--145.

\bibitem{shen2014modeling}
H.~Shen, D.~Wang, C.~Song, A.-L. Barab{\'a}si, Modeling and predicting
popularity dynamics via reinforced poisson processes, in: Proceedings of the
AAAI Conference on Artificial Intelligence, Vol.~28, 2014.

\bibitem{bao2015modeling}
P.~Bao, H.-W. Shen, X.~Jin, X.-Q. Cheng, Modeling and predicting popularity
dynamics of microblogs using self-excited hawkes processes, in: Proceedings
of the 24th International Conference on World Wide Web, 2015, pp. 9--10.

\bibitem{lymperopoulos2021rc}
I.~N. Lymperopoulos, Rc-tweet: modeling and predicting the popularity of tweets
through the dynamics of a capacitor, Expert Systems with Applications 163
(2021) 113785.

\bibitem{foroozani2019anomalous}
A.~Foroozani, M.~Ebrahimi, Anomalous information diffusion in social networks:
Twitter and digg, Expert Systems with Applications 134 (2019) 249--266.

\bibitem{bourigault2014learning}
S.~Bourigault, C.~Lagnier, S.~Lamprier, L.~Denoyer, P.~Gallinari, Learning
social network embeddings for predicting information diffusion, in:
Proceedings of the 7th ACM international conference on Web search and data
mining, 2014, pp. 393--402.

\bibitem{bourigault2016representation}
S.~Bourigault, S.~Lamprier, P.~Gallinari, Representation learning for
information diffusion through social networks: an embedded cascade model, in:
Proceedings of the Ninth ACM international conference on Web Search and Data
Mining, 2016, pp. 573--582.

\bibitem{liu2016learning}
W.~Liu, H.~Shen, W.~Ouyang, G.~Fu, L.~Zha, X.~Cheng, Learning cost-effective
social embedding for cascade prediction, in: Chinese National Conference on
Social Media Processing, Springer, 2016, pp. 1--13.

\bibitem{xie2021independent}
W.~Xie, X.~Wang, T.~Jia, Independent asymmetric embedding model for cascade
prediction on social network, arXiv preprint arXiv:2105.08291.

\bibitem{perozzi2014deepwalk}
B.~Perozzi, R.~Al-Rfou, S.~Skiena, Deepwalk: Online learning of social
representations, in: Proceedings of the 20th ACM SIGKDD international
conference on Knowledge discovery and data mining, 2014, pp. 701--710.

\bibitem{chung2014empirical}
J.~Chung, C.~Gulcehre, K.~Cho, Y.~Bengio, Empirical evaluation of gated
recurrent neural networks on sequence modeling, arXiv preprint
arXiv:1412.3555.

\bibitem{cao2020popularity}
Q.~Cao, H.~Shen, J.~Gao, B.~Wei, X.~Cheng, Popularity prediction on social
platforms with coupled graph neural networks, in: Proceedings of the 13th
International Conference on Web Search and Data Mining, 2020, pp. 70--78.

\bibitem{zhao2020deep}
Y.~Zhao, N.~Yang, T.~Lin, S.~Y. Philip, Deep collaborative embedding for
information cascade prediction, Knowledge-Based Systems 193 (2020) 105502.

\bibitem{xu2020casgcn}
Z.~Xu, M.~Qian, X.~Huang, J.~Meng, Casgcn: Predicting future cascade growth
based on information diffusion graph, arXiv preprint arXiv:2009.05152.

\bibitem{wang2021ccasgnn}
Y.~Wang, X.~Wang, T.~Jia, Ccasgnn: Collaborative cascade prediction based on
graph neural networks, arXiv preprint arXiv:2112.03644.

\bibitem{feng2021aecasn}
X.~Feng, Q.~Zhao, Y.~Li, Aecasn: An information cascade predictor by learning
the structural representation of the whole cascade network with autoencoder,
Expert Systems with Applications (2021) 116260.

\bibitem{sabour2017dynamic}
S.~Sabour, N.~Frosst, G.~E. Hinton, Dynamic routing between capsules, arXiv
preprint arXiv:1710.09829.

\bibitem{barabasi1999emergence}
A.-L. Barab{\'a}si, R.~Albert, Emergence of scaling in random networks, science
286~(5439) (1999) 509--512.

\bibitem{zhou2020continual}
F.~Zhou, X.~Jing, X.~Xu, T.~Zhong, G.~Trajcevski, J.~Wu, Continual information
cascade learning, in: GLOBECOM 2020-2020 IEEE Global Communications
Conference, IEEE, 2020, pp. 1--6.

\bibitem{zhang2016structure}
B.~Zhang, Z.~Qian, S.~Lu, Structure pattern analysis and cascade prediction in
social networks, in: Joint European Conference on Machine Learning and
Knowledge Discovery in Databases, Springer, 2016, pp. 524--539.

\bibitem{yano2010s}
T.~Yano, N.~Smith, What’s worthy of comment? content and comment volume in
political blogs, in: Proceedings of the International AAAI Conference on Web
and Social Media, Vol.~4, 2010.

\bibitem{grover2016node2vec}
A.~Grover, J.~Leskovec, node2vec: Scalable feature learning for networks, in:
Proceedings of the 22nd ACM SIGKDD international conference on Knowledge
discovery and data mining, 2016, pp. 855--864.

\bibitem{vaswani2017attention}
A.~Vaswani, N.~Shazeer, N.~Parmar, J.~Uszkoreit, L.~Jones, A.~N. Gomez,
L.~Kaiser, I.~Polosukhin, Attention is all you need, in: NIPS, 2017.

\bibitem{tan2020syntactic}
Y.~Tan, X.~Wang, T.~Jia, From syntactic structure to semantic relationship:
hypernym extraction from definitions by recurrent neural networks using the
part of speech information, in: International Semantic Web Conference,
Springer, 2020, pp. 529--546.

\bibitem{wang2017topological}
J.~Wang, V.~W. Zheng, Z.~Liu, K.~C.-C. Chang, Topological recurrent neural
network for diffusion prediction, in: 2017 IEEE International Conference on
Data Mining (ICDM), IEEE, 2017, pp. 475--484.

\bibitem{sun2013understanding}
L.~Sun, K.~W. Axhausen, D.-H. Lee, X.~Huang, Understanding metropolitan
patterns of daily encounters, Proceedings of the National Academy of Sciences
110~(34) (2013) 13774--13779.

\bibitem{yan2017universal}
X.-Y. Yan, W.-X. Wang, Z.-Y. Gao, Y.-C. Lai, Universal model of individual and
population mobility on diverse spatial scales, Nature communications 8~(1)
(2017) 1--9.

\bibitem{jia2017quantifying}
T.~Jia, D.~Wang, B.~K. Szymanski, Quantifying patterns of research-interest
evolution, Nature Human Behaviour 1~(4) (2017) 1--7.

\bibitem{zeng2019increasing}
A.~Zeng, Z.~Shen, J.~Zhou, Y.~Fan, Z.~Di, Y.~Wang, H.~E. Stanley, S.~Havlin,
Increasing trend of scientists to switch between topics, Nature
communications 10~(1) (2019) 1--11.

\end{thebibliography}

\end{document}